\documentstyle[12pt]{article}
\begin{document}
{\pagestyle{empty}
~~\\

\vskip 2.0cm
{\renewcommand{\thefootnote}{\fnsymbol{footnote}}
\centerline{\large \bf Parallel Tempering Algorithm for}

\vskip 0.5cm
\centerline{\large \bf Conformational Studies of Biological Molecules} 
}
\vskip 2.0cm
 
\centerline{Ulrich H.E.~Hansmann
\footnote{\ \ e-mail: hansmann@ims.ac.jp}} 
\vskip 1.5cm
\centerline {{\it Department of Theoretical Studies, 
Institute for Molecular Science}} 
\centerline {{\it Okazaki, Aichi 444, Japan}}

\medbreak
\vskip 3.5cm
 
\centerline{\bf ABSTRACT}
\vskip 0.3cm
   
The effectiveness of a new algorithm,
parallel tempering, is studied for numerical simulations of biological
molecules.  These molecules suffer from a rough energy landscape.
The resulting slowing down in numerical simulations is overcome by the new
method. This is demonstrated by performing simulations with high statistics 
for one of 
the simplest peptides, Met-enkephalin. The numerical effectiveness of the 
new technique  was found to be much better than traditional methods and
is comparable to sophisticated  
methods like generalized ensemble techniques. 
\vfill
\newpage}
 \baselineskip=0.8cm
\noindent
{\bf 1.~INTRODUCTION} \\
One way of investigating  biological molecules  is by means of computer
experiments. However, such numerical simulations can be 
notoriously difficult when the molecule is described by ``realistic''
energy functions where  interactions between all atoms are taken into
account. For a large class of molecules (for instance, peptides or proteins)  
the various competing  
interactions yield to frustration and a  rough energy landscape. Hence, at low
temperatures simulations based on 
canonical Monte Carlo \cite{Metro} or molecular dynamics techniques will 
get trapped in one of the  multitude of local minima separated
by high energy barriers. Only small parts of conformational space 
are sampled and physical quantities cannot be calculated accurately. 
One way  to overcome this difficulty is to perform a simulation in a so-called
{\it generalized ensemble}, which is based on a non-Boltzmann
probability distribution.
Multicanonical algorithms \cite{MU} and simulated tempering \cite{L} 
are prominent examples of such an approach which also includes the new
ensemble introduced in Refs.~\cite{H97a,HO96g}.
Application of  these techniques to the protein folding  problem was first 
addressed in Ref.~\cite{HO} and over the last few years the generalized 
ensemble approach has become 
an often used method for simulation of biological molecules and other
complex systems \cite{HO}-\cite{EH96d}.
A numerical comparison of  different generalized ensemble
algorithms can be found in Ref.~\cite{HO96b}.

Another way to alleviate the multiple minima problem 
is to look for improved updates of configurations
in the numerical simulation.
The cluster algorithm \cite{CA} is an example of global updates
that enhances thermalization and has
been very successful in spin systems.
However, for most other systems with frustration,
no such updates are known.
Here, a new algorithm, parallel tempering,  is described which 
introduces improved updates by means of constructing a special
generalized ensemble. It is shown that the 
 new method can be successfully applied to the simulation of molecules
with complex energy landscape. 
By simulating one of the  simplest peptides, Met-enkephalin, with high 
statistics  its effectiveness is compared with canonical simulations and
sophisticated generalized ensemble techniques like the multicanonical algorithm
\cite{MU}. Both Monte Carlo and molecular dynamics versions of
parallel tempering are studied and it is shown  that the new method   
 can be combined with other generalized ensemble techniques.\\  

\noindent
{\bf 2.~METHODS:}\\
Let me start by briefly summarizing 
the parallel tempering algorithm which was  originally introduced 
in the context of spin glass simulations \cite{PT3}. Similar ideas were
also proposed in Refs.~\cite{PT1,PT2}.

In a regular canonical simulation,  a configuration of {\it one} copy of 
the molecule under consideration is updated by Monte Carlo  \cite{Metro} 
or molecular dynamics. Each configuration $C$ is characterized by a set of 
(generalized) coordinates and is assigned a Boltzmann weight 
\begin{equation}
w_B (\beta,C) = e^{-\beta E(C)}~,
\end{equation}
where we have introduced the inverse temperature $\beta = 1/k_BT$ with $k_B$
the Boltzmann constant.  The so realized Markov - chain yields to a
canonical distribution at {\it one} temperature $T$. 

On the other hand, in {\it parallel tempering} one considers 
an artificial system build up of {\it N non--interacting}
copies of the molecule, each at a different temperature $T_i$. A state of 
the artificial system is specified by ${\bf \cal{C}} = \{C_1,C_2,...,C_N\}$,
where  each $C_i$  is a set of (generalized) coordinates which describes 
the configuration of the $i-$th copy. Since the $N$ copies of the molecule 
are not interacting with each other, one can assign to a state ${\bf \cal{C}}$
of the compound system a weight: 
\begin{equation}
w_{PT} ({\bf \cal{C}}) = e^{\displaystyle -\sum_i^N \beta_i E(C_i)} 
                        = \prod_{i}^{N} w_{B} (\beta_i,E(C_i))~.
\label{wcs}
\end{equation}
Without lack of generality one can 
assume $\beta_1<\beta_2< .... <\beta_N$. For a numerical simulation of 
the artificial system one has to construct a Markov-chain which will ensure that
the corresponding equilibrium distribution will be approached. 
Such a Markov-chain can be realized with the following  two sets of
moves \cite{PT3}:
\begin{enumerate}
\item Standard MC  moves which effect only one, 
      say the $i$-th, copy.  These moves are called {\it local} updates
      because they change only one coordinate of the configuration in
      solely one copy. Since the copies are non-interacting it follows from
      Eq.~\ref{wcs} that the transition probability depends only on the
      change in potential energy of the $i-$th copy. Hence, such local MC moves
      are accepted or rejected according to the usual 
      Metropolis criterion \cite{Metro} with probability:
      \begin{equation}
      w_{PT}({\bf \cal{C}}^{old} \rightarrow {\bf \cal{C}}^{new}) 
      = w_{B}(C_i^{old} \rightarrow C_i^{new}) 
      = \min(1,e^{-\beta_i \Delta_i E})~,
      \label{wcan}
      \end{equation}
      where $\Delta_i E$ is defined as 
      $\Delta_i E = E(C_i^{new}) - E(C_i^{old})$.
\item Exchange of conformations between two  copies $i$ and $j=i+1$:
      \begin{eqnarray}
      C_{i}^{new} & = & C_{j}^{old}\\
      C_{j}^{new} & = & C_{i}^{old}~.
      \end{eqnarray}
      Such exchange is a {\it global} update in the sense that for
      the $i-$th copy the whole configuration changes (and the same for
      the $j-$th copy). Since this move introduces configurational 
      changes in {\it two} copies of the molecule, it follows from 
      Eq.~\ref{wcs} that the  exchange is accepted or rejected according  
      to the Metropolis criterion with probability: 
      \begin{eqnarray}
        w({\bf \cal{C}}^{old} \rightarrow {\bf \cal{C}}^{new}) & = & \min(1, 
        e^{-\beta_i E(C_j) - \beta_j E(C_i) + \beta_i E(C_i) +\beta_j E(C_j)})
       \\
       & = & \min(1,e^{(\beta_j - \beta_i)(E(C_j)-E(C_i))})\\
       & = & \min(1,e^{\Delta \beta \Delta E})\\
       & = & \min(1,e^{\Delta})~,
      \label{wexch}
      \end{eqnarray}
where  $\Delta = \Delta\beta \Delta E$, $\Delta \beta = \beta_j - \beta_i$ 
and $\Delta E = E(C_j) - E(C_i)$.
While it is not necessary  to restrict the
exchange  to pairs of copies associated with neighboring inverse 
temperatures $\beta_i$ and $\beta_{i+1}$, this choice will be optimal,
since the acceptance ratio will decrease exponentially with the difference
$\Delta\beta = \beta_j - \beta_i$. 
\end{enumerate}

It is interesting to observe that from the view point of the artificial
compound system 
 the above set of moves realizes a {\it generalized ensemble} simulation 
with (non-canonical) weights given by Eq.~\ref{wcs}. However, one
can also choose the point of view that parallel tempering realizes
for each of the copies a {\it canonical} simulation at 
corresponding temperature $T_i$. The exchange of conformations  is then a
new and improved update  which decreases
the correlation between configurations (for each copy or temperature)
and hence increases thermalization of the {\it canonical} simulation
for each copy (temperature).
This guarantees in turn that each of the copies will approach its
equilibrium distribution (i.e. the Boltzmann distribution at temperature 
 $T_i$) much faster than without that global update.  From this point
of view it is also  obvious
that expectation values of a physical quantity $A$ are calculated
as usual according to:
\begin{equation}
 <A>_{T_i} = \frac{1}{MES}\sum_k^{MES} A(C_i(k))~,
\end{equation}
where $MES$ is the number of measurements taken for the $i$-th copy. 
Using re-weighting  
techniques \cite{FS} it is also possible to calculate expectation
values for intermediate 
inverse temperatures $\beta$ with $\beta_i < \beta < \beta_{i+1}$.

While parallel tempering is not restricted to the use on parallel computers, 
it is obvious that the new technique is well suited for them. A parallel
implementation can be easily realized by setting each of the copies on a
different node. On each node the molecule is simulated 
simultaneously and independent from the other nodes  with  Boltzmann
weights $e^{-\beta_i E(C_i)}$  by common  Monte Carlo 
techniques. After a few MC steps  pairs of configurations 
   $C_i$ and $C_j$ are exchanged between the  nodes $i$ and $j$ with a
   probability given by Eq.~\ref{wexch}. 
While the exchange of conformations has to be done by 
a Monte Carlo procedure, it is not necessary to use Monte Carlo for 
the {\it local} updates of the conformations on each node. 
Instead one can also evaluate on each node 
 for some time $\tau$  a molecular dynamics trajectory, using   one of the  
common canonical molecular dynamics techniques, before an exchange of 
conformations between two nodes is tried. It is also possible to 
replace the  exchange of configurations 
between different nodes  by an  exchange of temperatures between 
nodes. For a parallel implementation of the algorithm this  has 
the advantage that 
less messages have to be passed between the different nodes (two
temperature values instead of two sets of coordinates). 

The main advantage of parallel tempering over  generalized ensemble
methods is that the weights are {\it a priori} known, since the weight
for a state of the artificial system of non-interacting copies (see
Eq.~\ref{wcs}) is solely the product of
the Boltzmann weights for each of the copies. However, to ensure that an
exchange of either conformation or temperatures between two copies  will
happen with sufficient probability the differences in (inverse) temperatures 
$\Delta \beta$ in Eq.~\ref{wexch} have to be small.  With 
$\beta_j = \beta_{i+1} = \beta_i + \Delta \beta$ and approximating the
energy $E(C_{i})$ ($E(C_{i+1})$) by the thermal expectation value 
$<E>_{\beta_i}$ ($<E>_{\beta_{i+1}}$) one can write the
logarithm of probability $e^{-\Delta}$ of an exchange in Eq.~\ref{wexch}  as
\begin{equation}
\Delta \approx (\Delta \beta)^2 \frac{d}{d\beta} <E>~.
\label{cond}
\end{equation}
It is obvious from this equation that parallel tempering is not suitable for
simulations of first order phase transitions, since in that case
$<E>$ is not continuous at the critical temperature $T_c$ (and the
distribution of energies $P(E)$ bimodal).  For all other cases it follows
 from the above equation that an exchange of configurations
between two copies will happen with sufficient probability, 
as long as $\Delta$ is of order of 
one. Since the average energy  grows roughly 
proportional with the number of residues $N$, $\Delta \beta$ should
be of order of $1/\sqrt N$ to satisfy that condition. Hence, the number of
temperatures to simulate should also increase roughly 
proportional to $\sqrt N$ with
the number of residues. However, the problem remains of finding the
number and distribution of temperatures which guarantees optimal 
performance of the algorithm.  The lowest temperature will depend on the
molecule under consideration and physical quantities one is interested, but it
will in general be a temperature where usual canonical simulations  
get trapped in local minima. In the present algorithm, escape from these
minima is possible through the exchange of conformation between two copies
associated with different temperatures. Hence, the highest temperature
has to be chosen such that any energy barrier can be crossed at this
temperature. In this way it is guaranteed that by the successive 
exchange of conformations between copies 
any energy barrier can be overcome  and all of the replica will thermalize.

It is easy to combine parallel tempering with other
generalized ensemble techniques, since  the algorithm  only requires
that the copies are non interacting and therefore the weight of the compound
system  factorizable into a
product of weights for each copy. Hence, one can generalize
Eq.~\ref{wcs} to
\begin{equation}
w_{PT} ({\cal{C}}) = \prod_{i=1}^{N} w_{GE} (f(C_i))~,
\label{wptnew}
\end{equation}
where $w_{GE} (f(C_i))$ is a generalized ensemble weight for the $i$-th copy.
The  modifications in the transition probabilities 
Eq.~\ref{wcan} and \ref{wexch} which follow from this generalization are
 straight forward. Performing a parallel tempering simulation with
generalized ensemble weights for the copies will yield to corresponding
non-canonical distributions. Hence, to calculate thermodynamic quantities
at temperatures $T_i$ one has to use re-weighting techniques \cite{FS}.

In the present article one examples of such a combination of 
parallel tempering with generalized ensemble techniques is studied. For this 
the ensemble of Refs.~\cite{H97a,HO96g} (which is closely related to
Tsalis generalized mechanics formalism \cite{Tsa}) was chosen and 
to some of the copies  the following  weight 
\begin{equation}
w_{GE}(E(\beta_i,C_i)) = \left(1+\beta_i\frac{E(C_i) - E_{0}}{m}\right)^{-m}
\label{gewei}
\end{equation}
was assigned. Here, $E_0$ is an
estimate for the ground state energy of the molecule and $m$ a free
parameter. Obviously, the new weight reduces in the low-energy  
region to the canonical Boltzmann weight $\exp (- \beta E)$ for 
$\frac{\beta (E-E_{0})}{m} \ll 1$. 
 On the other hand, high-energy regions are no
longer exponentially suppressed but only according to a power law,
which enhances excursions to high-energy regions. 
It is expected that the so-defined weights  increase
the probability of an exchange of configurations between copies.\\

\noindent
{\bf 3.~RESULTS AND DISCUSSION}\\
The effectiveness of the new simulation technique was tested for 
 Met-enkephalin, one of the simplest peptides, which has become  a often
used model to examine new algorithms. 
Met-enkephalin has the amino-acid sequence Tyr-Gly-Gly-Phe-Met.
The potential energy function
$E_{tot}$ that was used is given by the sum of
the electrostatic term $E_{es}$, 12-6 Lennard-Jones term $E_{vdW}$, and
hydrogen-bond term $E_{hb}$ for all pairs of atoms in the peptide together with
the torsion term $E_{tors}$ for all torsion angles:
\begin{eqnarray}
E_{tot} & = & E_{es} + E_{vdW} + E_{hb} + E_{tors},\\
E_{es}  & = & \sum_{(i,j)} \frac{332q_i q_j}{\epsilon r_{ij}},\\
E_{vdW} & = & \sum_{(i,j)} \left( \frac{A_{ij}}{r^{12}_{ij}}
                                - \frac{B_{ij}}{r^6_{ij}} \right),\\
E_{hb}  & = & \sum_{(i,j)} \left( \frac{C_{ij}}{r^{12}_{ij}}
                                - \frac{D_{ij}}{r^{10}_{ij}} \right),\\
E_{tors}& = & \sum_l U_l \left( 1 \pm \cos (n_l \chi_l ) \right),
\end{eqnarray}
where $r_{ij}$ is the distance between the atoms $i$ and $j$,
and $\chi_l$ is the $l$-th torsion angle.
The parameters ($q_i,A_{ij},B_{ij},C_{ij},
D_{ij},U_l$ and $n_l$) for the energy function were adopted
from ECEPP/2.\cite{EC}
The computer code SMC\footnote{The program SMC was written by
Dr.~Frank Eisenmenger (eisenmenger@rz.hu-berlin.de)}  was used. 
The peptide-bond
dihedral angles $\omega$ were fixed at the value 180$^\circ$
for simplicity,
which leaves 19 angles $\phi_i,~\psi_i$, and $\chi_i$ as
independent variables.

Parallel tempering simulations with 7 copies were performed. The
corresponding temperatures were $T_1 =1000$ K, $T_2 = 500$ K, $T_3 =
330$ K, $T_4 =250$ K, $T_5=170 $ K, $T_6=100$ K and $T_7 = 50$ K.
The simulation consists of 144,000 sweeps for each copy where in every sweep 
each of the 19 angles is updated once.  After  one sweep for each
copy   an exchange of conformations between pairs of copies at 
neighboring temperatures was tried simulatanously for each of the three pairs. 
 Hence, the total number of updates in the parallel tempering simulation was 
 $(19 + 3) \times 7 \times 144,000 = 22,176,000$. This large number
was chosen to ensure high statistics and  is similar to the 
1,000,000 sweeps ($=19,000,000 $ updates)  used in earlier work where also
Met-enkephalin was used to compare  the effectiveness of various 
numerical algorithms \cite{HO96b}. 
For the above number of updates 15 hour CPU time were needed on a
Silicon Graphics Indigo 2 workstation. While parallel tempering 
is best suited for parallel computers the workstation was used 
because it was easier accessible than the parallel machine at IMS 
computer center. 

The results of the parallel tempering simulation was compared with that of a 
multicanonical simulation  and canonical simulations keeping the
number of updates constant (preliminary runs had shown that all methods
need roughly the same amount of CPU time for a fixed number of updates). 
In the case of the multicanonical simulation this number includes 
the 100,000 sweeps
(1,900,000 MC updates) necessary to obtain the weights by the
iterative procedure described in Ref.~\cite{HO94c}. The canonical simulations 
were performed at the same 7 temperatures as used in the parallel
tempering method and each temperature was simulated with same number of updates
(3,168,000 MC updates). 

To compare the performance of the various algorithms 
the potential energy  and the overlap with the (known) ground
state was measured. The latter quantity indicates how much a given configuration
differs from the ground state and is given by
\begin{equation}
O(t) = 1 -\frac{1}{90~n_F} \sum_{i=1}^{n_F} |\alpha_i^{(t)}-
\alpha_i^{(GS)}|~,
\label{eqol}
\end{equation}
where $\alpha_i^{(t)}$ and $\alpha_i^{(GS)}$ (in degrees) stand for 
the $n_F$ dihedral angles of the conformation at $t$-th Monte Carlo
sweep 
and the ground-state conformation, respectively. Symmetries
for the side-chain angles were taken into account and the difference
$\alpha_i^{(t)}- \alpha_i^{(GS)}$ was always projected into the interval
$[-180^{\circ},180^{\circ}]$. The above  definition  guarantees that one has
\begin{equation}
0 \le ~<O>_T~ \le 1~,
\end{equation}
with the limiting values
\begin{equation}
\left\{
\begin{array}{rl}
 <O(t)>_T~~ \rightarrow 1~,~~&T \rightarrow 0~, \\
 <O(t)>_T~~ \rightarrow 0~,~~&T \rightarrow \infty~.
\end{array}
\right.
\end{equation}
 \ \\

I start presenting my results by showing in Fig.~1 the histogram
of energies as obtained from the parallel tempering simulation with
Monte Carlo updates.  Note the   overlap between the Boltzmann
distributions which correspond to neighboring temperatures. The energies
 where the two histograms have a common overlap are the ones where  the
transition probabilities of Eq.~\ref{wexch} are large enough to allow 
for an exchange of configurations.  The larger the overlap of
histograms the higher is the probability for an exchange of
conformations. Hence, the temperatures $T_i$ of the copies have to be 
chosen such that there is sufficient overlab between  Boltzmann 
distributions which correspond to neighboring temperatures. 

To demonstrate the time evolution of states in the 
parallel tempering algorithm  the  start configurations of each of 
the seven copies were marked by
a label. The labeled configurations were called ``replicas'' and their
evolution through the whole parallel tempering simulation was followed.  
Note that the ``replicas'' are independent from each other. Through
 the exchange moves of parallel tempering the seven ``replicas''  
are shuffled between the seven temperatures (copies), however, these
moves do not introduce any correlation between them. In Fig.~2a it is 
shown for one of the  ``replicas''  how the seven different temperatures
are visited in the course of the simulation. Due to this random walk in 
temperatures the replica changes dramatically between ground state like
conformers and random coils. This can be  seen from the 
corresponding ``time series'' of the overlap of the studied  replica with the
ground state  in Fig.~2b. The overlap varies through the simulation between
$O << 1$ $=$ (disordered states) and $O \approx 1$ (ordered state).
Note that the values of overlap in Fig.~2b and temperature in Fig.~2a are
correlated. Ground state like conformations ($O$ close to $1$) were
observed when the ``replica'' stayed at low temperatures, while  disordered 
structures (small values of $O$) appeared together with high temperatures.

To visualize the exchange of conformations by
parallel tempering for a single copy (temperature), the label of the 
``replicas'' which is visiting the copy associated with the lowest temperature  
($T=50$ K), is displayed in Fig.~3 as a function of simulation time (in
MC sweeps). Since the ``replicas''  are independent from each other, each
change of the ``replica''-label indicates that the new conformation is
no longer correlated with the previous conformation. Hence, through the 
exchange of conformations the Markov chain  converges much faster
to the stationary distribution than it does in the case of a regular canonical
simulation with only local Monte Carlo updates. This can be seen in
Fig.~4a where the ``time series'' in energy is displayed  for both a 
regular canonical simulation at $T=50$ K and for the  copy with $T=50$ K of a 
parallel tempering simulation. 
Obviously the regular canonical Monte Carlo got stucked in a
local  minimum and was not able to thermalize. From previous simulations
(see Ref.~\cite{HO96g}) it is known that even 1,000,000 sweeps are not
enough to thermalize Met-enkephalin at $T=50$ K. On the other hand,
with the exchange of configurations by  parallel tempering the simulation 
thermalizes at that temperature in less than 10,000 sweeps. This follows
also from Fig.~4b where  the  ``time series'' of the overlap function is 
displayed for both simulations. In the case of  parallel tempering 
at $T=50 K$ most of the conformations are close to the ground state
which is consistent with  observations from previous simulations
\cite{HO,HO94c}. On the other hand the regular canonical simulation got 
stucked in a conformation far from the ground state showing again
that the regular Monte Carlo simulation never converged to the true 
distribution. 

Hence, ignoring in the parallel tempering simulation the first  10,000 sweeps 
necessary for thermalization one can  calculate expectation values for
average energy and the overlap with the (known) ground state. The
obtained values were in all cases within the errorbars identical with
the ones obtained by  the multicanonical simulation  and agree with the results
of earlier work \cite{HOE96,EH96d}). On the other hand, the regular
canonical simulations  yield at low temperatures to different and
unreasonable values, since for these  temperatures the simulations never 
thermalized. The  data for the two  quantities are summarized in 
Tab.~1 and 2 for all three methods. 

An important question is how the effectiveness of new approach  compares
 with other methods and under which conditions it is optimal. To answer 
this question, further parallel tempering simulations with same 
number of updates were performed,  but where  either the  local 
Monte Carlo updates in Eq.~\ref{wcan} were replaced by  
molecular dynamics trajectories 
or the canonical weights for the copies in Eq.~\ref{wcs} by the the generalized 
ensemble weights of Eq.~\ref{gewei}.  The molecular   
dynamic updates were performed in dihedral space. The trajectories 
followed for each copy consisted of  19 leap frog steps with
time step $\Delta t= 0.005$ (in arbitrary units).  As an example 
for a combination of parallel tempering 
with generalized ensemble techniques, a simulation was done where
for the 3 lowest temperatures ($T=50,100$ and $170$ K) the weights of
Eq.~\ref{gewei} were used and Boltzmann weights for the higher temperatures.
The generalized ensemble weights for the three lowest temperatures
were chosen such that the resulting deviation from a canonical
distribution would be small. This was ensured by choosing $m=5 \times n_F=95$ 
(with $n_F$  the number of degrees of freedom) and an arbitrary value 
$E_0=-20.72$ kcal/mol.

In each 
case  expectation values for physical quantities were obtained which were
within the errorbars the same as the one given in Tab.~1 and 2. However,
the transition probabilities varied. It did not depend on whether Monte  
Carlo or molecular dynamics techniques were  used as local updates, but
only on the  chosen weights. This can be seen in Tab.~3 where these 
transition probabilities are summarized. As one can see the
probabilities for exchange of conformations between copies   
can be enhanced by choosing suitable non-canonical weights
like the ones defined in Eq.~\ref{gewei} and used here. 
To measure the effectiveness of parallel tempering  and  to compare it
with other sophisticated methods like  the multicanonical algorithm 
the number of ``tunneling events''  was measured. This quantity gives 
a lower bound for the
number of independent ground state conformers found in the simulation.
A tunneling event is defined as a random walk between a ground
state like conformer (defined by the condition that the overlap $O \ge 0.9$ and
that the potential energy is not more than 1 kcal/mol above the ground state
energy $E_{GS} = -10.72$kcal/mol) and  a coil conformer ($O \le 0.3$). 
Using Monte Carlo updates, only two tunneling events  (at $T=250$ K) 
were observed for all seven  canonical simulations  but 22 events in the case 
of parallel tempering. This number should be compared with that of
a multicanonical run of same statistics where  26 tunneling
events were found. Hence, one can conclude that both parallel tempering
and multicanonical algorithms are of similar efficiency and much better than
regular canonical Monte Carlo with solely local updates. 
The effectiveness of parallel tempering can be improved by
choosing suitable generalized ensemble weights. With the generalized ensemble
weights used here the number of tunneling events could be increased 
by a factor 1.5 to 34 events (reflecting the increased transition
probabilities betweeen the copies, see Tab.~3). However, a drawback of
such a combination  with generalized ensemble methods
is that the weights are no longer {\it a priori} given for parallel
tempering. The gained  improvement
requires careful choice of additional parameters and may not always be
worth the additional effort. For some other combinations 
of the exponent $m$ and the constant $E_0$ in Eq.~\ref{gewei} it was
found that the efficiency 
became even worse than for the case  where  canonical weights were
assigned to  all copies (data not shown). Hence, by assigning generalized 
ensemble weights to all or only some copies, the effectiveness of 
parallel tempering can be increased, but whether such approach is useful
or not, may depend on the molecule under investigation. 

\noindent
{\bf 4.~Conclusions}\\
It was   shown that a new algorithm, parallel tempering, can be
successful applied to simulation of molecules and helps
to overcome the multiple minima problem. The new method can be used with
both Monte Carlo and molecular dynamics updates.  Its effectiveness
is comparable to generalized ensemble techniques like the
multicanonical algorithm. However, unlike for the case of
generalized ensemble techniques,  the weights are {\it a priori} known
for parallel tempering, which makes application of the new method
technically easier than that of generalized ensembe algorithms.
Further, the two techniques can be combined which allows to increase 
their efficiency. 

\vspace{0.5cm} 
\noindent
{\bf Acknowledgements:}\\ 
 The simulation were performed on the computers at the Computer
Center at the Institute for Molecular Science (IMS), Okazaki,
Japan. The article was written when I was fellow at the Center for
Interdisciplinary Research (ZiF) of the Bielefeld University. I like
to thank  ZiF and specially F. Karsch, head of the research group 
``Multiscale Phenomena and their Simulation'',  
 for the kind hospitality extended to me.\\

\newpage
\noindent
{\Large Tables:}
\begin{itemize}
\item {Tab.~1: Average potential energy $<E>$ as function of
                  temperature.}
\begin{center}
\begin{tabular}{cccc}
 $T$ & Canonical & Multicanonical & Parallel Tempering\\ \hline
1000 & 17.48 (1)      &    17.50 (2)        &    17.51 (6)     \\
 500 &  8.65 (6)      &     8.65 (5)        &     8.75 (8)     \\
 330 &  1.44 (22)     &     1.64 (12)       &     1.56 (11)    \\
 250 & -3.23 (76)     &    -2.85 (15)       &    -2.84 (16)    \\
 170 & -4.27 (18)     &    -6.36 (7)        &    -6.27 (4)     \\
 100 & -1.08 (9)      &    -8.50 (5)        &    -8.49 (3)     \\
  50 & -2.54 (5)      &    -9.69 (3)        &    -9.68 (2)     \\
\end{tabular}
\end{center}
\  \\
\item{Tab.~2: Average overlap function (defined in Eq.~19) as
                  function of temperature.}
\begin{center}
\begin{tabular}{cccc}
 $T$ & Canonical & Multicanonical & Parallel Tempering\\ \hline
1000 &  0.29 (1)    &    0.29 (1)       &    0.29 (1)          \\
 500 &  0.34 (1)    &    0.34 (1)       &    0.34 (1)          \\
 330 &  0.48 (3)    &    0.45 (1)       &    0.44 (2)          \\
 250 &  0.62 (2)    &    0.61 (1)       &    0.60 (1)          \\
 170 &  0.10 (1)    &    0.78 (1)       &    0.78 (1)          \\
 100 &  0.34 (1)    &    0.89 (1)       &    0.89 (1)          \\
  50 &  0.42 (1)    &    0.94 (1)       &    0.94 (1)          \\
\end{tabular}
\end{center}
\  \\
\item{Tab.~3: Probability for an exchange of configuration between
         two temperatures 
         for various variants of the parallel tempering method.} 
\begin{center}
\begin{tabular}{cccc}\hline
  & Monte Carlo & Molecular Dynamics & Monte Carlo \\ \hline 
& Canonical weights & Canonical weights & Generalized Ensemble weights\\ \hline
$500~K \leftrightarrow 1000~K$   & 0.14& 0.14 &  0.14 \\
$330~k \leftrightarrow  500~K$   & 0.17& 0.18 &  0.18 \\
$250~k \leftrightarrow  330~K$   & 0.31& 0.28 &  0.29 \\
$170~k \leftrightarrow  250~K$   & 0.22& 0.29 &  0.74 \\
$100~k \leftrightarrow  170~K$   & 0.15& 0.16 &  0.21 \\
$ 50~k \leftrightarrow  100~K$   & 0.10& 0.10 &  0.31 \\
\end{tabular}
\end{center}
\end{itemize}

\newpage
\noindent
{\Large Figure Captions:}\\
\begin{itemize}
\item Fig.~1: Histogram of Energies for different temperatures as 
              obtained from  a parallel tempering simulation with
              7 copies and 144,000 Monte Carlo sweeps for each copy.
\item Fig.~2a: ``Time series'' of temperatures which one of the seven
              ``replicas'' encountered over the 144,000 Monte Carlo sweeps
              in the parallel tempering simulation.
\item Fig.~2b: ``Time series'' of the overlap function, defined in Eq.~19,
              for one of the seven ``replicas'' over 144,000 Monte Carlo
              sweeps in the parallel tempering simulation.
\item Fig.~3: ``Time series'' of ``replicas'' over 144,000 Monte Carlo
              sweeps in the parallel tempering simulation 
              as encountered for $T=50$ K. 
\item Fig.~4a: ``Time series'' of energy for $T=50$ K over 144,000
              Monte Carlo sweeps as obtained from the parallel
              tempering  algorithm and a regular canonical simulation.
\item Fig.~4b: ``Time series'' of the overlap function (defined in Eq.~19)
              for $T=50$ K over 144,000 Monte Carlo sweeps as obtained
              from the parallel tempering algorithm  and a regular
              canonical simulation.
\end{itemize}

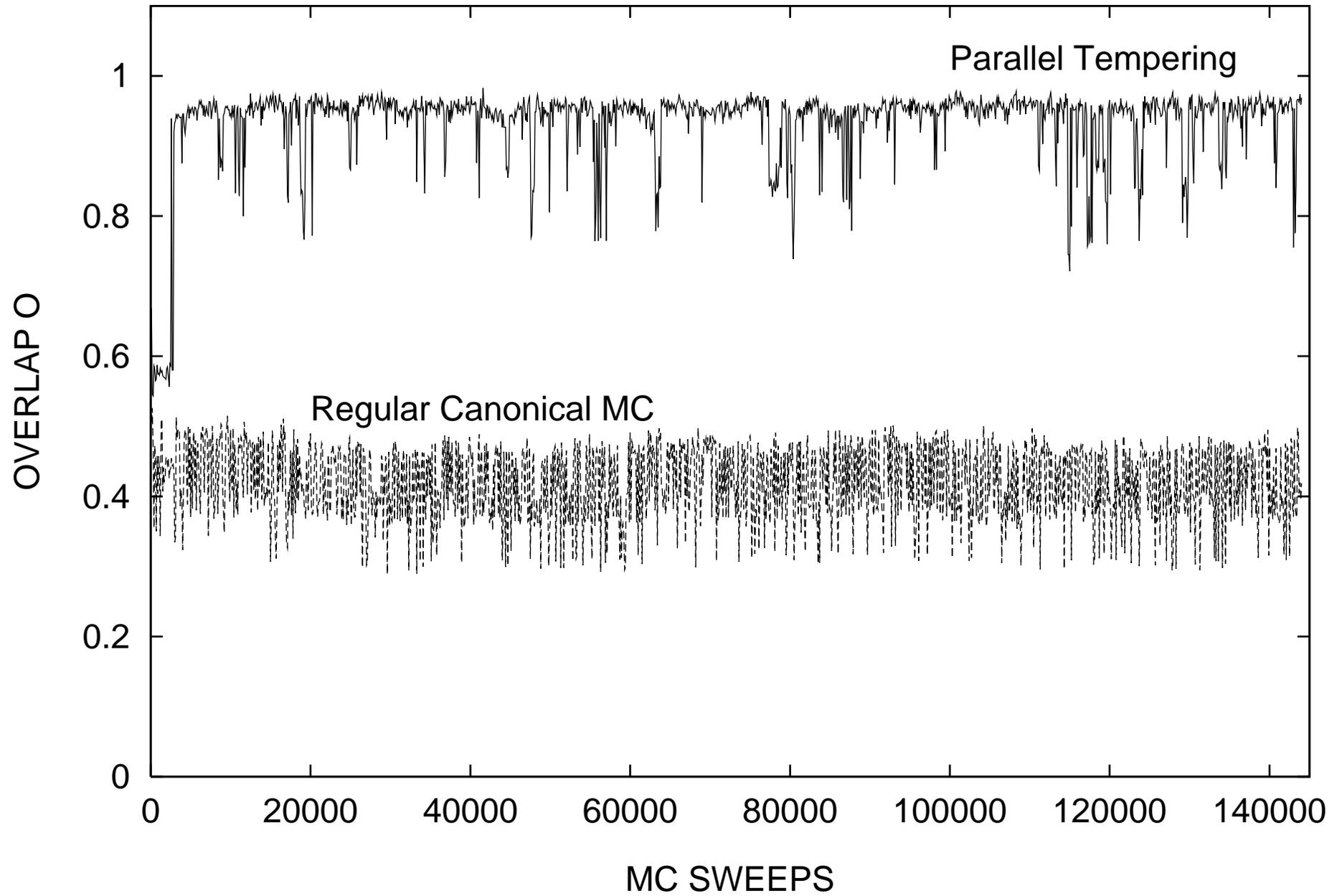
\end{document}